\documentclass[aps,prl,twocolumn,showpacs,floatfix]{revtex4-1}

\usepackage{amssymb}
\usepackage{amsmath,bm}
\usepackage[normalem]{ulem}
\usepackage[dvips]{color}
\usepackage{booktabs}
\usepackage{graphicx}
\usepackage{float}
\usepackage{multirow}
\usepackage{lineno}

\usepackage[normalem]{ulem}

\usepackage[usenames, dvipsnames]{xcolor}
\usepackage{tensor}

\newcommand{\xxi}{\rm $\Xi^-$}
\newcommand{\la}{\rm $\Lambda$}
\newcommand{\ka}{\rm $K^+$}
\newcommand{\pphi}{\rm $\phi$}
\newcommand{\snn}{$\sqrt{s_{\rm NN}}$}
\newcommand{\sratio}{$\frac{N(K^+)N(\Xi^-)}{N(\phi)N(\Lambda)}$}
\newcommand{\ds}{$\Delta s = \langle(\delta s)^2\rangle/\langle s\rangle^2$}

\renewcommand{\sout}{\bgroup \color{red} \ULdepth=-.5ex \ULset}

\begin{document}

\title{Probing QCD critical fluctuations from the yield ratio of strange hadrons in relativistic heavy-ion collisions}

\author{Tianhao Shao}\affiliation{Shanghai Institute of Applied Physics, Chinese Academy of Science, Shanghai 201800, China}\affiliation{University of Chinese Academy of Science, Beijing 100049, China}
\author{Jinhui Chen}\thanks{chenjinhui@fudan.edu.cn}\affiliation{Key Laboratory of Nuclear Physics and Ion-beam Application (MOE), Institute of Modern Physics, Fudan University, Shanghai 200433, China}
\author{Che Ming Ko}\thanks{ko@comp.tamu.edu}\affiliation{Cyclotron Institute and Department of Physics and Astronomy, Texas A\&M University, College Station, Texas 77843, USA}
\author{Kai-Jia Sun}\thanks{sunkaijiaxn@gmail.com}\affiliation{Cyclotron Institute and Department of Physics and Astronomy, Texas A\&M University, College Station, Texas 77843, USA}

\date{\today}
\begin{abstract}
By analyzing the available data on strange hadrons in central Pb+Pb collisions from the NA49 Collaboration at the Super Proton Synchrotron (SPS) and in central Au+Au collisions from the STAR Collaboration at the Relativistic Heavy-Ion Collider (RHIC) in a wide collision energy range from \snn~= 6.3 GeV to 200 GeV, we find a possible non-monotonic behavior in the ratio $\mathcal{O}_\text{K-$\Xi$-\pphi-\la}$= \sratio~of \ka, \xxi, \pphi, and \la~yields as a function of~\snn. Based on the quark coalescence model, which can take into account the effect of quark density fluctuations on hadron production, a possible non-monotonic behavior in the dependence of the strange quark density fluctuation on \snn~is obtained. This is in contrast to the coalescence model that does not include quark density fluctuations and also to the statistical hadronization model as both fail to describe even qualitatively the collision energy dependence of the ratio $\mathcal{O}_\text{K-$\Xi$-\pphi-\la}$. Our findings thus suggest that the signal and location of a possible critical endpoint in the QCD phase diagram, which is expected to result in large quark density fluctuations, can be found in the on-going Bean Energy Scan program at RHIC.
\end{abstract}


\maketitle


The main goal of the experiments on heavy-ion collisions at relativistic energies is to study the properties of strongly interacting matter under extreme conditions, especially those of the quark-gluon plasma (QGP), the hadronic matter, and the transition between these two phases of matter~\cite{RevModPhys.89.035001,BRAUNMUNZINGER201676,CHEN20181}. Studies based on lattice quantum chromodynamics (LQCD) calculations~\cite{Fodor:2004nz} and various effective models~\cite{Asakawa:1989bq,Stephanov:1998dy,Hatta:2002sj} have indicated that the transition between the QGP and the hadronic matter is a smooth crossover at vanishing baryon chemical potential ($\mu_B$) but likely changes to a first-order phase transition at large $\mu_B$, with an associated critical endpoint (CEP) or a tricritical endpoint seperating these two transitions~\cite{Ding:2015ona}. Locating the position of the CEP in the QCD phase diagram is one of the most important issues in particle and nuclear physics. To search for this CEP, experiments have been carried out already at the Beam Energy Scan programs at SPS~\cite{Kaon1Afanasiev:2002mx,Kaon2Alt:2007aa,phiAlt:2008iv,LaXi-SPSAlt:2008qm} and at RHIC~\cite{Aggarwal:2010cw,Luo:2017faz} as well as planned at the future Facility for Antiproton and Ion Research (FAIR) and Nuclotron-based Ion Collider Facility (NICA).

As suggested in Ref.~\cite{Asakawa:2000wh}, the QCD phase transition can be probed by studying the fluctuations of physical observables in relativistic heavy ion collisions. This is because enhanced long-wavelength fluctuations near the CEP can lead to singularities in all thermodynamic observables. In heavy-ion collisions, these fluctuations have been studied by using experimental data on an event-by-event basis and looking at event-by-event fluctuations~\cite{Friman2011,Endrodi:2011gv}. For example, the fourth-order fluctuation of net-proton distribution had been measured in the BES program by the STAR Collaboration, and a possible non-monotonic behavior in its dependence on the center-of-mass collision energy \snn ~was observed~\cite{Adamczyk:2013dal}. Also, large baryon density fluctuations are expected to be developed in the produced QGP when its evolution trajectory in the QCD phase diagram passes across the CEP~\cite{Asakawa:2000wh,Hatta:2003wn}. Recent studies based on both the hydrodynamic approach~\cite{Steinheimer:2012gc,Steinheimer:2013xxa} and the transport model~\cite{Li:2016uvu} have shown that the spinodal instabilities associated with a first-order QGP to hadronic matter phase transition at finite baryon chemical potential can generate appreciable fluctuations in the baryon density distribution. The CEP in the QCD phase diagram can thus be located from relativistic heavy ion collisions by studying the collision energy dependence of quark density fluctuations and determining the temperature and baryon chemical potential at the collision energy near which the quark density fluctuations show a non-monotonic behavior. This idea was used in Refs.~\cite{Sun:2017xrx,SUN2018499} to study the neutron density fluctuation in heavy ion collisions at the SPS energies in the framework of the nucleon coalescence model for light nuclei production, and a non-monotonic dependence on \snn~was found in the yield ratio $\mathcal{O}_\text{p-d-t}=N_{\rm p}N_{\rm ^3H}/N_{\rm d}^2$ of proton ($\rm p$), deuteron ($\rm d$), and triton ($\rm ^3H$). Although it was suggested in Refs.~\cite{Sun:2017xrx,SUN2018499} that the extracted collision energy dependence of neutron density fluctuation may originate from the light quark density fluctuations when the evolution trajectory of produced QGP passes through the CEP in the QCD phase diagram, a quantitative study of this relation based on a viable dynamic model is still missing.

To probe more directly the quark density fluctuations, it was suggested in Ref.~\cite{Ko2018} to study the yield ratio $\mathcal{O}_\text{K-$\Xi$-\pphi-\la}$ = \sratio~of \ka, \xxi, \pphi, and \la~in relativistic heavy ion collisions. This is because strange hadrons are known to scatter less frequently than nucleons during the hadronic evolution, and their final abundances at kinetic freeze out are expected to be similar to those at hadronization if both include the contribution from resonance decays. A non-monotonic behavior in the \snn~dependence of this ratio then indicates a possible strange quark density fluctuation in these collisions.

The use of strange hadrons produced in relativistic heavy ion collisions to probe the properties of QGP has a long history. Because the mass of strange quark has the same magnitude as the QGP phase transition temperature, strange quarks can be abundantly produced in QGP~\cite{Rafelski1982} and converted to strange hadrons after hadronization. Enhanced production of strange hadrons has thus been considered a good signature for the formation of QGP in relativistic heavy ion collisions. For example, the well known peak in the $\langle K^+ \rangle$/$\langle \pi^+ \rangle$ ratio in central Pb+Pb collisions at a beam energy of 30 A GeV~\cite{Kaon2Alt:2007aa}, the change of the $\Omega/\phi$ ratios scaled by the number of constituent quarks in central Au+Au collision between \snn~=11.5 GeV and \snn~$\geq$ 19.6 GeV~\cite{OmPhiAdamczyk:2015lvo}, the non-monotonic suppression of the nuclear modification factor $R_{\rm CP}$ for charged hadrons including the kaons from \snn~=62.4 GeV to 7.7 GeV~\cite{Adamczyk:2017nof} have all been considered as the signals for the onset of deconfinement transition in the matter produced in relativistic heavy ion collisions.

In the present study, we analyze the published data on \xxi, \ka, \la~and \pphi~yields in central Pb+Pb collisions at SPS energies from the NA49 Collaboration~\cite{Kaon1Afanasiev:2002mx,Kaon2Alt:2007aa,phiAlt:2008iv,LaXi-SPSAlt:2008qm} and in central Au+Au collisions at RHIC energies from the STAR collaboration~\cite{La-STARAdam:2019koz,K-STARAdamczyk:2017iwn,Phi-STARAdamczyk:2015lvo,STAR200,phi200,K130,la130,Xi130,Abelev:2008ab} to find the dependence of the ratio $\mathcal{O}_\text{K-$\Xi$-\pphi-\la}$ on \snn.  We then use the quark coalescence model~\cite{CSERNAI1986223,PhysRevC.66.025205,PhysRevLett.90.202303,PhysRevLett.90.202302,PhysRevLett.91.092301,PhysRevC.78.034907,PhysRevC.74.064902,Zhang:2019bkf}
to interpret the result. In particular, we show that in the quark coalescence model the ratio $\mathcal{O}_\text{K-$\Xi$-\pphi-\la}$ is sensitive to the strange quark relative density fluctuation \ds~at the QGP to hadronic matter phase transition. It is known from the success of the statistical model in describing the yield ratios of hadrons that the chemical freeze-out in heavy-ion collisions occurs at the phase transition temperature and remain essentially unchanged during the hadronic evolution. As shown in Ref.~\cite{Jun2017}, this is due to the constancy of entropy per particle during the evolution from the chemical to the kinetic freeze-out~\cite{Jun2017}. Because of the constancy of the yield ratios of hadrons during the hadronic evolution, studying their dependence on the collision energy is expected to provide a unique probe to the quark density fluctuations during the first-order phase transition of the QGP to the hadronic matter, which would help locate the CEP in the QCD phase diagram.

\begin{table*}[htbp]
\centering
\caption{Yields of $\Xi^-$, $K^+$, $\Lambda$ and $\phi$ in full rapidity space from central (0-7.2\% centrality) Pb+Pb collisions at SPS energies measured by the NA49 Collaboration~\cite{Kaon1Afanasiev:2002mx,Kaon2Alt:2007aa,phiAlt:2008iv,LaXi-SPSAlt:2008qm}. Only statistical uncertainties are listed. Also given are the yield ratio $\mathcal{O}_\text{K-$\Xi$-\pphi-\la}$ extracted from experimental data and the extracted relative strange quark fluctuation $\Delta s$ based on the quark coalescence model using the hadronization temperature $T_C$~\cite{Wheaton:2004qb}. The units for E, \snn,~and $T_C$ are A GeV, GeV, and MeV, respectively.}
\label{Tab1}

\begin{tabular}{lccccccccc}
\hline
E    & $\sqrt{s_{\rm NN}}$   & $\Xi^-$               & $K^+$             &$\Lambda$         &$\phi$         &$\mathcal{O}_\text{K-$\Xi$-\pphi-\la}$        &$T_C$  & $\Delta s$  \\
\hline \hline
20   &6.3                  &1.50$\pm$0.13                              &40.7$\pm$0.7                           &27.1$\pm$0.2                           &1.89$\pm$0.31     &1.19$\pm$0.22      &131.3     & 0.08 $\pm$ 0.20\\
30   &7.6                 &2.42$\pm$0.19                              &52.9$\pm$0.9                           &36.9$\pm$0.3                           &1.84$\pm$0.22     &1.88$\pm$0.27      &140.1    &0.71 $\pm$ 0.25\\
40   &8.8                 &2.96$\pm$0.20                              &59.1$\pm$1.9                           &43.1$\pm$0.4                           &2.55$\pm$0.17     &1.59$\pm$0.16       &146.1    & 0.44 $\pm$ 0.15\\
80   &12.3               &3.80$\pm$0.26                              &76.9$\pm$2.0                           &50.1$\pm$0.6                           &4.04$\pm$0.19     &1.44$\pm$0.13      &153.5      & 0.31 $\pm$ 0.12\\
\hline
\end{tabular}
\end{table*}

\begin{table*}[htbp]
\centering
\caption{Same as TABLE~\ref{Tab1} for midrapidity strange hadrons except the last two columns, which give the yield ratio $\mathcal{O}_\text{K-$\Xi$-\pphi-\la}$ from the statistical model~\cite{Wheaton:2004qb} and the coalescence model using the hadronization temperature $T_C$. For the statistical model, the percentage of contributions from different decay channels is taken from that calculated at \snn~= 200 GeV.}
\label{Tab2}
\begin{tabular}{lccccccccccc}
\hline
E    & $\sqrt{s_{\rm NN}}$    &$\Xi^-$           & $K^+$      &$\Lambda$    &$\phi$               &$\mathcal{O}_\text{K-$\Xi$-\pphi-\la}$       &$T_C$  &$\Delta s$       &stat. model    &COAL-SH   \\
\hline \hline
20   &6.3               &0.93$\pm$0.13    &16.4$\pm$0.6  &13.4$\pm$0.1 &1.17$\pm$0.23     &0.97$\pm$0.24   &131.3 &   0$^{+0.10}$  &1.30           &1.10       \\
30   &7.6               &1.17$\pm$0.13    &21.2$\pm$0.8  &14.7$\pm$0.2 &0.94$\pm$0.13     &1.79$\pm$0.33   &140.1 &    0.63 $\pm$ 0.30 &1.40           &1.10       \\
40   &8.8              &1.15$\pm$0.11    &20.1$\pm$0.3  &14.6$\pm$0.2 &1.16$\pm$0.16     &1.36$\pm$0.23   &146.1  &    0.24 $\pm$ 0.21 &1.33           &1.10       \\
80   &12.3           &1.22$\pm$0.14    &24.6$\pm$0.2  &12.9$\pm$0.2 &1.52$\pm$0.11     &1.53$\pm$0.21   &153.5    &    0.39 $\pm$ 0.19 &1.23           &1.10      \\
\hline
\end{tabular}
\end{table*}

\begin{table*}[htbp]
\centering
\caption{Same as Table~II for central Au+Au collisions at RHIC energies from the STAR Collaboration~\cite{La-STARAdam:2019koz,K-STARAdamczyk:2017iwn,Phi-STARAdamczyk:2015lvo,STAR200,phi200,K130,la130,Xi130,Abelev:2008ab} except the centrality is 0-5\% at $\sqrt{s}=200$ GeV and the quoted errors for $K^+$ are the quadratic sum of statistical and the dominant systematic uncertainties.}
\label{Tab3}
\begin{tabular}{lccccccccccc}
\hline
$\sqrt{s_{\rm NN}}$    & $\Xi^-$               &$K^+$          &$\Lambda$              &$\phi$             &$\mathcal{O}_\text{K-$\Xi$-\pphi-\la}$   &$T_C$(MeV)    & $\Delta s$           &stat. model   &COAL-SH   \\
\hline \hline
7.7                   &1.11$\pm$0.02     &19.06$\pm$0.42    &13.21$\pm$0.08     &1.23$\pm$0.11        &1.30$\pm$0.12        &144.3     &   0.18 $\pm$ 0.11    &1.40           &1.10       \\
11.5                   &1.21$\pm$0.01     &22.89$\pm$0.47    &12.62$\pm$0.06     &1.68$\pm$0.11        &1.31$\pm$0.09       &149.4    &   0.19 $\pm$ 0.08   &1.27           &1.10       \\
19.6                   &1.50$\pm$0.01     &26.94$\pm$0.53    &11.37$\pm$0.03     &2.58$\pm$0.14        &1.38$\pm$0.08       &153.9    &    0.25 $\pm$ 0.07  &1.21           &1.10       \\
27                    &1.49$\pm$0.01    &28.48$\pm$0.56    &10.65$\pm$0.03     &3.05$\pm$0.17        &1.31$\pm$0.08         &155.0     &    0.19 $\pm$ 0.07  &1.20           &1.10       \\
39                    &1.39$\pm$0.01    &29.88$\pm$0.58    &9.70$\pm$0.02     &3.33$\pm$0.17        &1.29$\pm$0.07          &156.4     &   0.17 $\pm$ 0.06    &1.20           &1.10       \\
130                   &2.04$\pm$0.16       &42.10$\pm$0.42    &15.00$\pm$0.27      &5.73$\pm$0.37        &1.00$\pm$0.10     & 165    & 0$^{+0}$      &1.20           &1.10       \\
200                    &2.17$\pm$0.06       &51.30$\pm$6.50      &14.80$\pm$0.20      &7.39$\pm$0.11        &1.02$\pm$0.13  &166  &  0$^{+0.05}$        &1.20           &1.10       \\
\hline
\end{tabular}
\end{table*}

\begin{figure}[!h]
\includegraphics[scale=0.43]{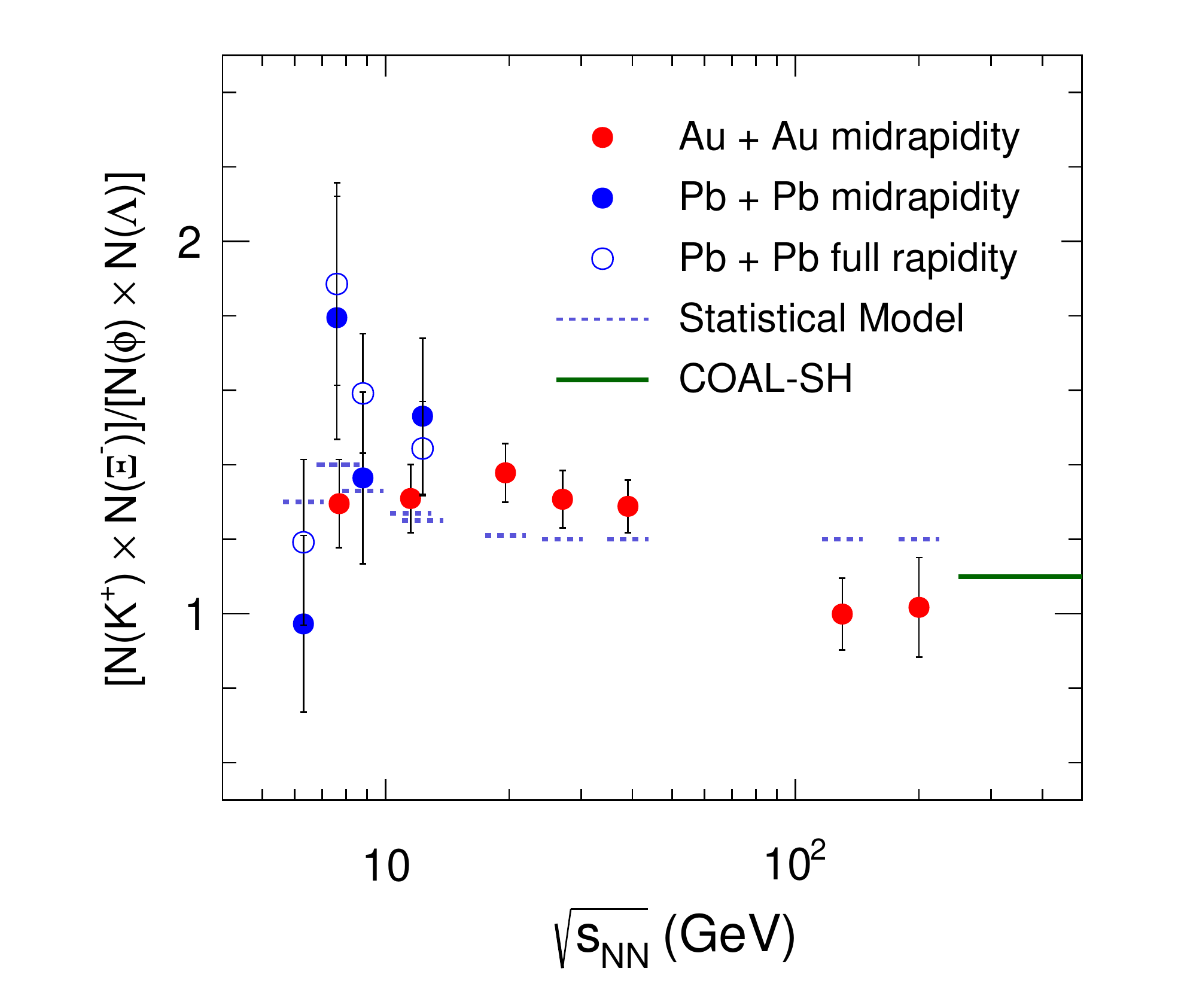}
\caption{Collision energy \snn~dependence of the ratio $\mathcal{O}_\text{K-$\Xi$-\pphi-\la}$=\sratio~in central Pb+Pb collisions at SPS energies and in central Au+Au collisions at RHIC energies. Filled and open circles denote, respectively, the ratio obtained from midrapidity and in full rapidity space. Error bars represent the statistical uncertainties. The horizontal line on the right side of the figure show the ratio calculated from the COAL-SH model without quark density fluctuations~\cite{Sun:2017ooe}. The dash lines are the ratio calculated from the statistical model~\cite{Wheaton:2004qb}.}
\label{Fig1}
\end{figure}

We first summarize in Tables~\ref{Tab1} and \ref{Tab2} the experimental data in full rapidity space and midrapidity, respectively, from Pb+Pb collisions and in Table~\ref{Tab3} those in midrapidity from Au+Au collisions. Also shown are the ratio $\mathcal{O}_\text{K-$\Xi$-\pphi-\la}$ extracted from these data. To see more clearly the collision energy dependence of the ratio, we plot Fig.~\ref{Fig1} its dependence on \snn, where only the statistical errors are shown because most of the systematic errors cancel out in calculating the ratio. As in the analysis based on the statistical hadronization model~\cite{PhysRevC.82.011901}, we do not include the data from Pb+Pb collisions at 158 A GeV because of the different centrality bins used for \ka, \pphi, \xxi, and \la.  A non-monotonic behavior in the dependence of the ratio $\mathcal{O}_\text{K-$\Xi$-\pphi-\la}$ on the collision energy is clearly seen at \snn~$\sim$~8 GeV, which is similar to that found in Refs.~\cite{Sun:2017xrx,SUN2018499} from the yield ratio $\mathcal{O}_{\text{p-d-t}}$.

To interpret the experimental results, we use the analytical coalescence formula COAL-SH developed in Ref.~\cite{Sun:2017ooe} to calculate the hadron yield at the QGP to hadronic matter phase transition temperature $T_C$. According to this formula, the yield $N_h$ of a hadron species $h$ consisting of $A$ constituent quarks of mass $m_q$ from a QGP of $N_q$ number of quark species $q$ uniformly distributed in a volume $V_C$ is given by
\begin{eqnarray}\label{Eq1}
N_h &=& g_cg_{\rm rel}g_{\rm size}\left(\sum_{i=1}^Am_i\right)^{3/2}\bigg[\prod_{i=1}^A \frac{N_i}{m_i^{3/2}}\bigg] \nonumber\\
  &\times& \prod_{i=1}^{A-1} \frac{(4\pi/\omega)^{3/2}}{V_Cx(1+x^2)}\bigg(\frac{x^2}{1+x^2}\bigg)^{l_i}G(l_i,x).
\end{eqnarray}
In the above, $g_c = (2S+1)/6^A$ is the coalescence factor for colored constituent quarks of spin $1/2$ to form a colorless hadron of spin S,  $g_{\rm size}$ is the correction due to the finite size of produced hadron, and  it is taken to be one due to the very large size of the QGP compared to the sizes of produced hadrons, and $g_{\rm rel}$ is the relativistic correction given by
\begin{eqnarray}\label{Eq1.1}
g_{\rm rel} \approx \left[1+\frac{15}{8}T\left(\sum_{i=1}^{A}m_i\right)^{-1}\right]\prod_{i=1}^A \left(1+\frac{15}{8}\frac{T}{m_i}\right)^{-1}.
\end{eqnarray}
For the quantities in the second line of the equation, $\omega$ is the oscillator frequency used to obtain the quark wave functions inside the hadron, $x = (2T_C/\omega)^{1/2}$, $l_i$ is the orbital angular momentum associated with the $i$-th relative coordinate, and $G(l,x)=\sum_{k=0}^l\frac{l!}{k!(l-k)!}\frac{1}{(2k+1)x^{2k}}$ is the suppression factor due to the orbital angular momentum on the coalescence probability.

For the four strange hadrons considered in the present study, their constituent quarks  are all in the $s$-state ($l$ = 0) according to the constituent quark model,  which leads to $G(l,x) = 1$. The yields for these four strange hadrons are then given by:
\begin{eqnarray}
\label{Eq2}
N_{K^+}& = & g_{K^+} \frac{(m_u + m_{\bar{s}})^{3/2}}{m_u^{3/2}m_{\bar{s}}^{3/2}} \frac{N_uN_{\bar{s}}}{V_C} 
\frac{(2\pi/T_C)^{3/2}}{1+\omega/(2T_C)},\\
\label{Eq3}N_{\Xi^-} &= & g_{\Xi^-} \frac{(m_d + 2m_s)^{3/2}}{m_d^{3/2}m_s^3} \frac{N_dN_s^2}{V_C^2}
\frac{(2\pi/T_C)^3}{[1+\omega/(2T_C)]^2},\\
\label{Eq4}N_{\phi} &=& g_{\phi} \frac{(m_s + m_{\bar{s}})^{3/2}}{m_s^{3/2}m_{\bar{s}}^{3/2}} \frac{N_sN_{\bar{s}}}{V_C} 
\frac{(2\pi/T_C)^{3/2}}{1+\omega/(2T_C)},\\
\label{Eq5}N_{\Lambda} &=& g_{\Lambda} \frac{(m_u + m_d + m_s)^{3/2}}{m_u^{3/2}m_d^{3/2}m_s^{3/2}} \frac{N_uN_dN_s}{V_C^2} 
\frac{(2\pi/T_C)^3}{[1+\omega/(2T_C)]^2},\nonumber\\
\end{eqnarray}
where $g_{K^+}=g_{\rm rel, K^+}/36$, $g_{\Xi^-}=g_{\rm rel, \Xi^-}/108$, $g_{\phi}=g_{\rm rel, \phi}/12$, and $g_{\Lambda}=g_{\rm rel, \Lambda}/108$.

The above results can be generalized to take into account quark density fluctuations by expressing the quark density distributions as
\begin{eqnarray}\label{Eq6}
n_q(\vec{r}) = \frac{1}{V_C}\int n_q(\vec{r})d\vec{r} + \delta n_q(\vec{r}) = \langle q\rangle + \delta q(\vec{r}),
\end{eqnarray}
where $\langle \cdot \rangle$ denotes the average value over the coordinate space and $\delta q(\vec{r})$ with $\langle \delta q\rangle=0$ is its deviation from the average value $\langle q\rangle$. Defining the quark relative density fluctuations $\Delta q=\langle(\delta q)^2\rangle/\langle q\rangle^2$ and the quark density fluctuation correlation coefficients $\alpha_{q_1q_2}=\langle\delta q_1\delta q_2\rangle/(\langle q_1\rangle\langle q_2\rangle)$, and neglecting higher-order correlation coefficients $\alpha_{q_1q_2q_3}=\langle\delta q_1\delta q_2\delta q_3\rangle/(\langle q_1\rangle\langle q_2\rangle\langle q_3\rangle)$, Eqs.~(\ref{Eq2})-(\ref{Eq5}) can be rewritten as

\begin{eqnarray}
\label{Eq17}N_{K^+} &=& g_{K^+}\frac{(2\pi/T_C)^{3/2}}{1+\omega/(2T_C)}V_C\langle \bar{s}\rangle\langle u\rangle(1 + \alpha_{\bar{s}u}),\\
\label{Eq18}N_{\Xi^-} &=& g_{\Xi^-}\frac{(2\pi/T_C)^3}{[1+\omega/(2T_C)]^2}
V_C\langle s\rangle^2\langle d\rangle\nonumber\\
&&\times(1 + \Delta s + 2\alpha_{sd}),\\
\label{Eq19}N_{\phi} &=& g_{\phi}\frac{(2\pi/T_C)^{3/2}}{1+\omega/(2T_C)}V_C \langle s\rangle\langle \bar s\rangle(1 + \alpha_{s\bar{s}}),\\
\label{Eq20}N_{\Lambda} &=& g_{\Lambda}\frac{(2\pi/T_C)^3}{[1+\omega/(2T_C)]^2}
V_C\langle s\rangle\langle u\rangle\langle d\rangle\nonumber\\
&&\times(1 + \alpha_{sd} + \alpha_{su}+ \alpha_{ud}).
\end{eqnarray}
These equations then lead to the following expression for the ratio $\mathcal{O}_\text{K-$\Xi$-\pphi-\la}$:
\begin{eqnarray}\label{Eq21}
\mathcal{O}_\text{K-$\Xi$-\pphi-\la} &=& \frac{1}{3} \frac{g_{\rm rel,K^+}g_{\rm rel,\Xi^-}}{g_{\rm rel,\phi}g_{\rm rel,\Lambda}}\nonumber\\
&\times&\frac{(m_u + m_{\bar{s}})^{3/2}(m_d + 2m_s)^{3/2}}{(m_s + m_{\bar{s}})^{3/2}(m_u + m_d + m_s)^{3/2}} \nonumber\\
&\times& \frac{(1 + \alpha_{\bar{s}u})(1 + \Delta s+2\alpha_{sd})}{(1 + \alpha_{s\bar{s}})(1+\alpha_{sd} + \alpha_{su} + \alpha_{ud})}.
\end{eqnarray}

In the absence of quark density fluctuations, the yield ratio becomes simply $\mathcal{O}_\text{K-$\Xi$-\pphi-\la}=g$
with the constant $g$ denoting the expressions in the first two lines of Eq.(\ref{Eq21}). Since the value of $g$ changes very little for temperature between 100 and 160 MeV, the value of $
\mathcal{O}_\text{K-$\Xi$-\pphi-\la}$ is thus essentially independent of the collision energy, contradicting to the non-monotonic dependence on the collision energy seen in experiments. However, the value of $g$ depends on the contribution to the four ground state strange hadrons from decays of other
strange hadrons and their resonances that are included in the COAL-SH calculations. In the present study, 
we include the decay of $\Xi(1530)$ to \xxi, the decay of $\Sigma^0(1192)$, $\Sigma(1385)$, 
$\Lambda(1405)$ and $\Lambda(1520)$ to $\Lambda$, and the decay of $K^*(892)$, $K_1(1270)$ and $K_1(1400)$ 
to $K^+$. The resulting numbers of $K^+$, $\Xi^-$, and $\Lambda(1115)$ that should be compared with the measured ones are $N_{K^+}^{\text{measured}} = N_{K^+} + 
\frac{1}{3}N_{K^{*+}(892)} + \frac{2}{3}N_{K^{*0}(892)}+0.51N_{K_1^+(1270)} +0.47N_{K_1^0(1270)} 
+0.56N_{K_1^+(1400)}+0.44N_{K_1^0(1400)}\approx 10 N_{K^+}$, $N_{\Xi^-}^{\text{measured}} = N_{\Xi^-} + \frac{1}
{2}N_{\Xi(1530)} = 3N_{\Xi^-}$, and $N_{\Lambda}^{\text{measured}} = N_{\Lambda} + \frac{1}{3} N_{\Sigma(1192)} + 
(0.87+\frac{0.11}{3})N_{\Sigma(1385)} + \frac{1}{3}N_{\Lambda(1405)}+0.25N_{\Lambda(1520)}= 
8.27N_{\Lambda(1115)}$. In obtaining the contribution from $\Lambda(1405)$ and $\Lambda(1520)$ to $\Lambda(1115)$, we have taken into account the suppression due to quark orbital angular momentum as one of the quarks in these two resonances is in the $p-$state ($l=1$). For the $\phi$ meson, we assume no strong and electromagnetic decay corrections from resonances and thus have $N_{\phi}^{\text{measured}} = N_{\phi}$. Here $N_{K^+}$, $N_{\Xi^-}$, $N_{\phi}$ and $N_{\Lambda}$ represent the corresponding hadron yields obtained directly from the quark coalescence model. Using the constituent quark masses $m_u=m_d=0.3$ GeV and $m_s=0.5$ GeV, the phase transition temperature $T_C$ taken from the parametrization given in Ref.~\cite{Cleymans:2005xv} based on the statistical model fit to the available experimental data, we obtain the values of $g=1.1$ from COAL-SH. The corresponding value for the yield ratio $\mathcal{O}_\text{K-$\Xi$-\pphi-\la}=g$ is shown by the solid line on the right side of Fig.~\ref{Fig1} and also given in Tables~\ref{Tab2} and~\ref{Tab3}. It is seen that the yield ratio $\mathcal{O}_\text{K-$\Xi$-\pphi-\la}$ extracted from experimental data at \snn = 200 GeV, where one does not expect any quark density fluctuations, is described reasonably well by COAL-SH.

Eq.(\ref{Eq21}) shows that to extract the strange quark relative density fluctuation $\Delta s$, one needs information on the quark density fluctuation correlation coefficients $\alpha_{q_1q_2}$. Without information about these coefficients, we consider two extreme cases of uncorrelated and strongly correlated density fluctuations of quarks of different flavors. For uncorrelated quark density fluctuations, one has $\langle\delta q_1\delta q_2\rangle=\langle\delta q_1\rangle\langle\delta q_2\rangle$ and thus $\alpha_{q_1q_2}=0$, which leads to
\begin{eqnarray}\label{Eq22}
\mathcal{O}_\text{K-$\Xi$-\pphi-\la} = g(1 + \Delta s).
\end{eqnarray}
The yield ratio $\mathcal{O}_\text{K-$\Xi$-\pphi-\la}$ is then linearly proportional to the strange quark relative density fluctuation $\Delta s$. For strongly correlated quark density fluctuations, one has instead $\langle\delta q_1\delta q_2\rangle=\sqrt{\langle(\delta q_1)^2\rangle\langle(\delta q_2)^2\rangle}$ and thus $\alpha_{q_1q_2}=\sqrt{\Delta q_1\Delta q_2}$. The yield ratio is then
\begin{eqnarray}\label{Eq23}
\mathcal{O}_\text{K-$\Xi$-\pphi-\la}&=&g\frac{1 + \sqrt{\Delta \bar{s}\Delta u}}{1+\sqrt{\Delta s \Delta \bar{s}}}\nonumber\\
&\times&\frac{1 + \Delta s+2\sqrt{\Delta s\Delta d}}{1+\sqrt{\Delta s\Delta d} + \sqrt{\Delta s\Delta u} + \sqrt{\Delta u\Delta d}}.
\end{eqnarray}
In the limit of SU(3) symmetry, i.e., the $u$, $d$, and $s$ current quark masses are the same, one has $\Delta u=\Delta d=\Delta s$. The yield ratio in this case then has the constant value $\mathcal{O}_\text{K-$\Xi$-\pphi-\la}=g$ and is independent of the collision energy. Since the SU(3) symmetry is broken with the strange quark mass much larger than those of $u$ and $d$ quarks, which would lead to different interactions for $s$ quarks than $u$ and $d$ quarks in hot dense quark matter~\cite{Song:2012cd}, $\Delta s$ could be different from $\Delta u$ and $\Delta d$. As a result, one expects the yield ratio $\mathcal{O}_\text{K-$\Xi$-\pphi-\la}$ to show some collision energy dependence.

We note a similar expression can be derived for the dependence  of the yield ratio $\mathcal{O}_\text{$\bar K^0$-p-$\pi^+$-$\Lambda$}=\frac{N_{\bar K^0}N_p}{N_{\pi^+}N_\Lambda}$ on the $u$ quark relative density fluctuation, but it is a challenging task to measure $\bar K^0$ in experiments since it always mixes with $K^0$.

To illustrate the possible dependence of the strange quark relative density fluctuation on the collision energy in heavy ion collisions, we show in Tables \ref{Tab1}-\ref{Tab3} the extracted values of $\Delta s$ for the case of $\alpha_{q_1q_2}=0$ using the value $g=1.1$ from COAL-SH. It is seen that $\Delta s$ shows a non-monotonic behavior in its dependence on the collision energy, and this can be understood as follows. For central collisions at higher incident energies when the baryon chemical potential of produced QGP is small, the phase transition from QGP to hadronic matter is likely a smooth crossover. The density fluctuations in the produced matter at these collision energies is thus insignificant. As the collision energy decreases, the produced matter may have its evolution trajectory in the temperature and baryon chemical potential plane pass by or approach closely to the CEP of the QCD phase diagram and can thus develop a large density fluctuation. With further decrease in collision energy, its trajectory moves away from the CEP and enters the region of a first-order phase transition.  Because of the spinodal instability associated with the first-order phase transition~\cite{Steinheimer:2012gc}, the hot dense matter may also develop large density fluctuations. With further decrease in collision energy, the density fluctuation diminishes as a result of the smaller size and shorter lifetime of the produced QGP. In this picture, the non-monotonic behavior shown in Fig.~\ref{Fig1} for the \snn~dependence of the ratio $\mathcal{O}_\text{K-$\Xi$-\pphi-\la}$ thus indicates that the evolution trajectory of the produced QGP in these collisions may have reached or closely approached the CPE or have undergone a first-order phase transition.

The non-monotonic collision energy dependence of the ratio $\mathcal{O}_\text{K-$\Xi$-\pphi-\la}$, particularly the possible peak at \snn$\sim 8$ GeV cannot be explained by the statistical hadronization model~\cite{KOCH1986167,Cho:2017dcy,Cho:2010db,Cho:2011ew} either. In this model, the number of hadrons of a given type $h$ produced at the chemical freeze-out temperature $T_C$ and volume $V_C$ is given in the non-relativistic limit by
\begin{eqnarray}\label{Eq24}
N_{h}^{stat} = \gamma_{h} g_{h} V_C\bigg(\frac{m_hT_C}{2\pi}\bigg)^{3/2}e^{-m_{h}/T_C},
\end{eqnarray}
where $g_{h}$ is the degeneracy of the hadron, $\gamma_{h}$ is the fugacity, and $m_{h}$ is the mass of the hadron. The ratio $\mathcal{O}_\text{K-$\Xi$-\pphi-\la}$ in this model is then given by
\begin{eqnarray}\label{Eq25}
\mathcal{O}_\text{K-$\Xi$-\pphi-\la} &=& \frac{1}{3}\frac{m_{K^+}^{3/2} m_{\Xi^-}^{3/2}}{m_{\phi}^{3/2} m_{\Lambda}^{3/2}} \nonumber \\
&\times&  e^{-(m_{\Xi^-}+m_{K^+}-m_{\Lambda}-m_{\phi})/T_C}.
\end{eqnarray}
To include the contribution from resonance decays, we use the THERMUS package~\cite{Wheaton:2004qb} with the chemical freeze-out temperature and baryon chemical potential determined in Ref.~\cite{Cleymans:2005xv}. The obtained results for the ratio $\mathcal{O}_\text{K-$\Xi$-\pphi-\la}$ of midrapidity $K^+$, $\Xi^-$, $\phi$ and $\Lambda(1115)$ are summarized in Tables~\ref{Tab2} and \ref{Tab3} and also shown by the dashed lines in Fig.~\ref{Fig1}. We note that the value of the ratio $\mathcal{O}_\text{K-$\Xi$-\pphi-\la}$ in central (0-10\%) Pb+Pb collisions at LHC energy of 2.76 TeV is $1.09\pm0.07$, which is consistent with the value of 1.20 from the statistical model calculation. Results from both the COAL-SH model and the statistical model clearly fail to quantitatively describe the non-monotonic collision energy dependence of the ratio  $\mathcal{O}_\text{K-$\Xi$-\pphi-\la}$ extracted from the experimental data in a wide range of collision energies.

We note that the uncertainties of the extracted ratio $\mathcal{O}_\text{K-$\Xi$-\pphi-\la}$ from experimental data are large.  Further experimental and theoretical investigations are needed to verify and extend the present results and to eventually establish the strange hadron yield ratio as a robust probe to the QCD critical point. In this respect, the ongoing phase II of the BES program at RHIC, which can measure precisely the total multiplicity of these strange hadrons in full rapidity space, will be very useful for studying the CEP in the QCD phase diagram via the yield ratio of strange hadrons. It is also of interest to study the strange quark density fluctuation in small collision systems to compare with the results from large collision systems. The ALICE collaboration~\cite{ALICE:2017jyt} has recently measured the multiplicity dependence of the yield ratio of multistrange baryons in $pp$ collisions at $\sqrt{s}$ = 7 TeV, and it will be very useful to have data also for charged kaon and $\phi$ meson from events with small multiplicities.

In summary, we have found a possible non-monotonic behavior in the collision energy dependence of the yield ratio  $\mathcal{O}_\text{K-$\Xi$-\pphi-\la}$ that is extracted from measured yields of strange hadrons in central Pb+Pb collisions at SPS energies from the NA49 Collaboration and in central Au+Au collisions at RHIC energies from the STAR Collaboration. This behavior cannot be explained by the usual coalescence model with uniform quark density distributions and by the statistical hadronization model. Including quark density fluctuations in the coalescence model allows one to extract the collision energy dependence of the strange quark fluctuation from that of $\mathcal{O}_\text{K-$\Xi$-\pphi-\la}$. Our study thus suggests that the CEP in the QCD phase diagram may have been reached or closely approached in heavy ion collisions at SPS and at RHIC BES energies. Future studies of strange hadron production in the phase II of BES program at RHIC and other heavy-ion collisions experiments, which will provide more accurate measurements of their multiplicities, are needed to versify present observations and to further advance our knowledge on the location of the CEP in the QCD phase diagram.

The authors thank Xianglei Zhu for helpful discussions. The work of T.S. and J.C. was supported in part by the National Natural Science Foundation of China under Contract Nos. 11890710, 11775288, 11421505 and 11520101004, while that of C.M.K. and K.J.S. was supported by the US Department of Energy under Contract No. DE-SC0015266 and the Welch Foundation under Grant No. A-1358.

\bibliography{myref-b2}
\end{document}